# Measurement of Photocarrier Mean Free Path via Speckled Laser Pump - Transient Fourier Microscopy Probe

Zuni Luo[1,2], Yifan Zhang[1,2], Jie Wei[2], Zhikun Xie[2], Tianshu Lai[2], and Ke Chen[1,2, *]

[1]Center for Neutron Science and Technology, Guangdong Provincial Key Laboratory of Magnetoelectric Physics and Devices, School of Physics, Sun Yat-sen University, 510275 Guangzhou, China
[2] State Key Laboratory of Optoelectronic Materials and Technologies, School of Physics, Sun Yat-sen University, 510275 Guangzhou, China

**ABSTRACT**. The mean free path of photocarriers is a crucial parameter for material design, device optimization, and new optoelectronics applications. Currently, this parameter remains unknown for many materials, and experimental means available for its measurement are considerably lacking. Meanwhile, it remains an unclear issue whether the mean free path of the photogenerated high-energy "hot" carriers is significantly different from that of the local-equilibrium-state carriers near the Fermi surface or around the band edge. Based on the concept of transient grating Fourier transform and utilizing a virtual lock-in amplification technique, we proposed and demonstrated an efficient experimental technique for measuring the mean free path of photocarriers. This method has facilitated direct observation of the photocarrier transport behavior across the transition between diffusive and ballistic motion, from which we surprisingly find that the mean free path of photogenerated hot carriers in Silicon membrane and GaAs quantum well can reach micron scale, more than an order of magnitude larger compared to the electrically-measured one. This work provides new ideas for characterization of photoelectronic devices under operating status and is expected to greatly enhance the understanding of the photocarrier transport process in opto-electronic or photonic materials.

Photocarrier mean free path (PMFP) is one of the core parameters determining the performance of semiconductor optoelectronic devices. When the key dimensions of these devices (e.g. the thickness of the absorption layer and the width of the depletion region) exceed the PMFP, the carriers will undergo frequent scattering, resulting in energy relaxation or recombination, leading to a decrease in collection efficiency. Therefore, regulating the PMFP [1-4] is a significant research direction in material design, device optimization and new optoelectronics applications. In addition, for the photodetection schemes, the PMFP under electric fields determines the saturation drift velocity and restricts the response speed [2-6]. And in solar cells, the ballistic transport within PMFP of nonthermalized electrons due to the bulk photovoltaic effect directly affects the photoelectric conversion which can even break through the Shockley-Queisser efficiency limit [7-10].

Several reports have indicated that high-energy photocarriers in a non-equilibrium state exhibit significantly larger mobility [8,10] and diffusion coefficient [11-13], compared to those low-energy carriers near the Fermi level or band edge, even reaching two orders of magnitude, which implies that the photocarrier has a much longer mean free path (MFP). Moreover, researchers have pointed out that the MFP estimated from the velocity of the equilibrium carriers that are thermalized with the lattice should be treated as the lower limit of the hot carrier MFP, as the non-equilibrium hot carriers owning higher energies have higher temperature and higher velocity than the lattice-thermalized state [7]. However, theoretical consensus has not yet been reached on how MFP values compare across different electronic states. For example, when considering inelastic scattering, electrons at sufficiently high energy levels are expected to have a longer MFP [14,15]. While another theoretical work shows that the electrons in the bottom of conduction band exhibit a longer MFP, attributing to a long relaxation time due to small density of states [16]. Currently, all existing experimental means for measuring MFP are only capable of detecting carriers in a quasi-equilibrium steady state, and their general principles are extracting the carrier scattering time and the carrier thermal velocity or Fermi velocity [17-22]. Experimental determination of MFP in a non-equilibrium state is still an unmet technological challenge, which hinders deeper understanding of carrier transport physics.

Aim to measure non-equilibrium PMFP, we were inspired by the transient grating spectroscopy (TGS) [23], which has achieved direct measurement of the phonon MFP [24]. Due to the size effect [25], when half of the grating period (the equivalent transport length) gradually decreases and becomes comparable or shorter than PMFP, the carrier transport state will transition from diffusive to ballistic, manifesting different transport and signal behavior, hence we can directly extract PMFP from the evident transformation. However, affected by the diffraction limit, it is difficult for the grating period to approach a very tiny size in common optical experimental systems. Moreover, to observe the gradual trend departing from the diffusive transport behavior, one should repeatedly change the grating period, which greatly increases the time cost and the operational difficulty. Fortunately, we realized that all of the aforementioned problems can be effectively addressed by performing speckled laser pump - Fourier microscopy probe measurement.

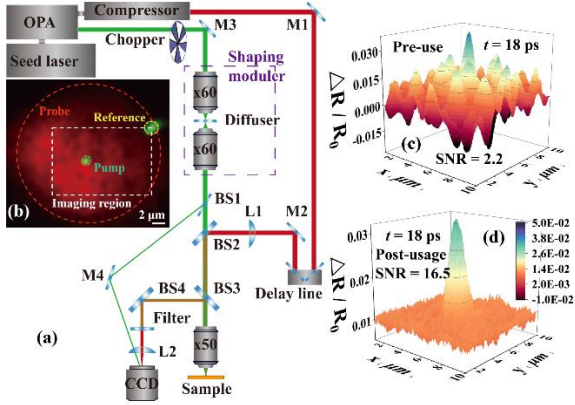

FIG 1. The schematic diagram of the experimental setup and result of the VLA. (a) The experimental optical path. A reference beam is composed of BS1 and M4, while a shaping module contains two high-magnification objectives and a diffuser. (b) The view of the camera. (c) and (d) show the reflection image signal $\Delta R/R_0$ at $t = 18\ ps$ when pre-use and post-usage the VLA. Clearly, the signal is almost submerged in the system noise before using. Owing to this technique, the SNR increases from 2.2 to 16.5, effectively extracting the accurate Gaussian profile of the carrier distribution.

In this paper, we propose a technique named transient Fourier microscopy (TFM) with the purpose of measuring non-equilibrium PMFP. The method extracts the characteristic decay time in the amplitude evolution under different wave vectors by performing Fourier transformation on the speckled laser pump-probe reflection image signal, and obtains the dynamic transport information of photoexcited hot carriers. Considering that TFM involves area detection demanding high fidelity in recording the detail of carrier spatial distribution, we incorporated with a virtual lock-in amplification technique (VLA) which demodulates the weak signals submerged in the noise by using their frequency and phase characteristics. With our proposed TFM method, we successfully and "sort of surprisingly" measured the PMFP of a $2\mu m$-thick Silicon membrane to be $\sim 0.65\mu m$, and that of a GaAs quantum well to be $\sim 1.25\mu m$. This work provides a new idea for gaining the PMFP experimentally and is expected to greatly improve the understanding of the carrier transport in many opto-electronic or photonic materials.

The concept of speckled laser pump - TFM is analogous to the TGS: gaining transport information by measuring the attenuation of the sinusoidal wave with wave vector $q$. Nevertheless, our TFM's advantage becomes apparent when considering that TGS can only measure the evolution of one specific sinusoidal wave generated by the interference of two pump pulses, corresponding to a single density grating, whereas speckled laser pump - TFM can simultaneously acquire multiple sinusoidal waves by simply performing Fourier analysis to the carrier distribution images at different time delays. Meanwhile, the tiny size of the speckled pump laser can enlarge the wave vector range, which means that the sizes of decomposed gratings become more widely distributed.

The experimental setup is formed by adding a reference light and a shaping module to the conventional ultrafast pump-probe reflectivity microscopy, as shown in Fig. 1(a). The VLA we proposed consists of signal input, reference acquisition and phase-sensitive detection, which can achieve ultra-narrowband demodulation under optical modulation frequency $f = 10\ Hz$. This mean can increase the signal-to-noise ratio (SNR) by more than seven times (see Fig. 1(c) and Fig. 1(d)) so as to accurately visualize the carrier distribution on an ultrafast time scale and effectively realize weak energy density pump measurement. The detailed introduction is in the Supplemental Material [26]. All the measurements are carried out at room temperature.

To verify the feasibility of TFM in principle, we firstly tested it with a numerical simulation. We generated a series of images which present photocarriers' random diffusion over time (see Fig. 2(a) and Fig. 2(b)), assuming that the initial distribution of energy carriers at $t = 0$ was a $\delta$ function at $(x, y) = (0,0)$, then the distribution at any later time $t$ would be a Gaussian function, $n(x, y, t) = exp(-t/\tau) \cdot 1/4\pi Dt \cdot exp(-(x^2 + y^2)/4Dt)$, where the preset diffusion coefficient $D = 1$ and carrier lifetime $\tau = 10^3$. Next performing Fourier transform on them to extract the evolution of the amplitude at different wave vectors $q$. The amplitude at each $q$ will decay exponentially, with the decay rate determined by the product of effective diffusion coefficient $D$ and the wave vector square $q^2$, as characterized in previous single-wave vector TGS experiment [24,28]:

$$|FFT.(n(x,y,t))| = Amp(q,t) \\ = A_0(q) \cdot e^{-t/\tau} \cdot e^{-Dq^2 t} + B(q) \quad (1)$$

where $A_0(q)$ is the amplitude of the reflection signal $\Delta R/R_0$ at the initial time, the term $e^{-t/\tau}$ describes the carrier relaxation after the photo-excitation, containing the decay information as a consequence of carrier recombination, the term $e^{-Dq^2 t}$ describes the smear out of the transient grating due to carrier transport from peak to valley, while $B(q)$ is the system background noise, which is independent of time. Fig. 2(c) shows the decrement of the normalized amplitude spectrum at the wave vector $q$ ranging from 0.1 to 0.4. Obviously, as the time $t$ increases, the amplitude gradually decreases, which is consistent with the spatial-temporal evolution of photocarriers. At the instance of the laser excitation, the carrier concentration injected into the pump region is the highest, then diffusing outward and simultaneously disappearing due to electron-hole recombination. Fourier transform just decomposes this diffusion-recombination process of Gaussian-shape carriers into decay processes of independent multiple sinusoidal transient gratings. According to the TGS [23] or Equ. (1), the characteristic grating decay time $\tau_q$ is given by

$$\frac{1}{\tau_q} = Dq^2 + \frac{1}{\tau} \quad (2)$$

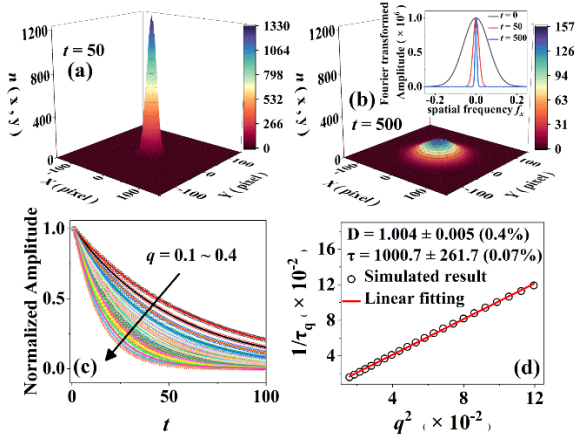

FIG 2. The numerical simulation results of TFM. (a) and (b) are the distribution of photocarriers at $t = 50$ and $t = 500$, respectively. The inlaid figure is amplitudes of the Fourier transform at different spatial frequency along $x$ axis versus time. (c) The evolution of the normalized amplitude spectrum under different wave vectors $q$. (d) The fitting results at $q = 0.1 \sim 0.4$.

We collect the characteristic times $\tau_q$ under the different wave vector $q$ in Fig. 2(d). The slope of linear fitting is diffusion coefficient $D$ $(1.004 \pm 0.005)$, while the reciprocal of the intercept yields carrier lifetime $\tau$ $(1000.7 \pm 261.7)$. The relative error is within 0.4%, which confirms the validity and the reliability of TFM.

According to Fourier transform, the effective range of the wave vector $q$ is related to the morphology of the photocarriers. As for the Gaussian beam excitation, the resolvable maximum wave vector $q_{max} = 2\pi/w$, where $w$ is the width of pump beam acting on the material. Theoretically, a sharp Gaussian pump beam with high intensity is desired, because it can offer not only a broad $q$ range but also large amplitudes of the photocarrier gratings. However, such high-intensity laser beam is prone to injuring the sample. Therefore, we present a better and practical method: applying random laser speckles which are simply generated by a diffuser and widely-distributed in the imaging area as the pump (see Fig. S1(a) and Fig. S1(b) [26]). In this way we utilized a larger surrounding area to generate similar amount of photocarrier as the desired sharp intensive Gaussian without sample damage. Owing to multi-beam interference and large angular spectral component, the size of the speckles, i.e. the sub-spots, can be commonly smaller than the minimum light-spot size directly formed by an on-axis low-angular-spectral Gaussian beam. Therefore, speckled laser pump means a larger $q_{max}$, thus is more likely to reach the ballistic transport region of photocarriers.

With the help of VLA, we use TFM to analyze the photocarrier transport in a $2\mu m$-thick Silicon membrane excited by a Gaussian beam and a reshaped speckled beam respectively. Fig. 3(a) and Fig. 3(b) respectively show the attenuation of the normalized amplitude spectrum at three different wave vectors $q$ and they all exhibit the expected exponentially decaying pattern. In Fig. 3(c), the red squares represent the decay rate of Fourier amplitude for Gaussian pump, while the black circles are those for the speckle pump. Since the speckles possess tiny sub-spots size and complex morphology comparing to Gaussian beam, its range of $q$ is wider, with $q_{max} = 6.7 \ \mu m^{-1}$. By fitting the linear diffusion region at $q = 1.1 \sim 4.8 \ \mu m^{-1}$, the bipolar diffusion coefficient $D_{G-Si} = 9.5 \pm 0.2 \ cm^2/s$ and the carrier lifetime $\tau_{G-Si} = 2.9 \pm 0.5 \ ns$ in the case of the Gaussian pump, while under the excitation of the speckle pump, $D_{S-Si} = 9.6 \pm 0.3 \ cm^2/s$ and $\tau_{S-Si} = 2.3 \pm 0.7 \ ns$. The same original data collected under the Gaussian pump is also analyzed by the transient reflectivity microscopy [13], the measured diffusion coefficient $D_{TRM-Si} = 9.5 \pm 0.4\ cm^2/s$, as shown in Fig. 3(d), which is consistent with that obtained by TFM. Besides, we have also performed multiple single-wave vector TGS experiments, which yields a consistent result (see Fig. S5 [26]). All our measured diffusion coefficients are close to the reported values [28,29]. Moreover, these measured carrier lifetimes also verge on $\tau_{PP-Si}$ ($2.6 \pm 0.3ns$) obtained by the pump-probe technique (see Fig. S6 (a) [26]). We have verified that the pump-probe experimental response is linear with pump fluence (see Fig. S7 (a) and Fig. S7 (b) [26]), which ensures that the carrier lifetime and the diffusion coefficient are constant within the excited carrier density range ($1.4 \sim 5.7 \times 10^{20}\ cm^{-3}$) in our experiments. All the above results successfully confirm the feasibility of TFM we proposed.

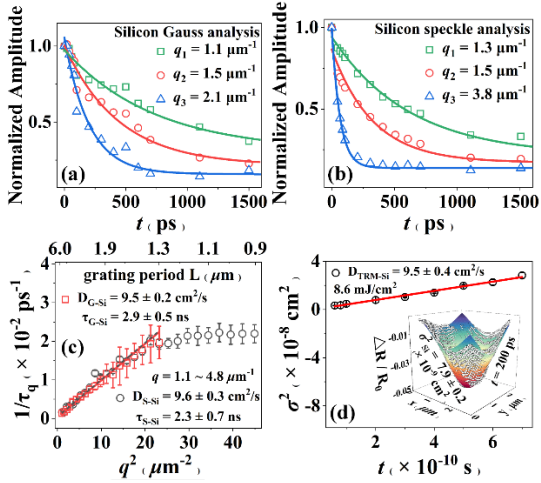

FIG 3. The experimental results of $2\mu m$-thick Silicon membrane with $515nm$ pump and $650nm$ probe. (a) and (b) respectively display the evolution of the normalized amplitude spectrum under Gaussian pump and speckle pump. (c) The experimental results obtained by TFM, where the speckle pump (black open circles) visualizes the process of carrier transport transitioning from diffusion to ballistic. (d) The result of transient reflectivity microscopy obtained from the same raw data under Gaussian pump. The inlaid figure is the fitting result of the 2D Gaussian profile at $t = 200\ ps$.

In Fig. 3(c), one wave vector $q$ corresponds to one carrier grating with a period of $L$ (upper axis). When half of the grating period ($L/2$) is less than the PMFP, the photocarriers will move from the peak of the grating to the neighbor valley directly, without participating any scattering. This ballistic kind of motion does not interact with the environment and maintains a low effective density flux, resulting in a decrease of the effective diffusion coefficient, known as the size effect [24]. From Fig. 3(c), it can be intuitively seen that as the period $L$ decreases, the carrier transport state does change from diffusive to ballistic, with the effective diffusion coefficient tending to decrease. And this transition i.e., the deviation from the diffusive slope, occurs at around $L_{cross} = 1.3\ \mu m$. According to quasi-ballistic electron transport theory, the effective carrier diffusion coefficient can be expressed as [26]:

$$D_{eff} = C \cdot \frac{l}{l+MFP} = \begin{cases} C \cdot l/MFP & (l \ll MFP) \\ C & (l \gg MFP) \end{cases} \quad (3)$$

where $C$ is the diffusion coefficient of diffusive regime, $l$ denotes transportation characteristic length. For our system, the $l$ corresponds to half a grating period because the source and drain of the carriers are equivalent to the peaks and valleys of the grating. In Fig. 4(a), we plot the effective diffusion coefficient $D_{eff}$ (the slope of the data in Fig. 3(c)) as a function of $l$, and obtain the clear "size-effect" behavior of $D_{eff}$, well consistent with Eq. (3). The transition point shown in Fig.4(a) indicates that the temporally-and-modally average photocarrier MFP of silicon $PMFP_{Si}$ is ~ $0.65\mu m$ ($L_{cross}/2$). We realized that $0.65\mu m$, measured for photocarriers under the bipolar transport condition where the transport characteristics are predominantly governed by the slower moving carrier (hole in silicon), appears to be an astonishingly huge value for carrier MFP from the conventional "local equilibrium steady state" perspective. Indeed, we employed four-probe method to measure the resistivity of our silicon sample, and by adopting reference mobilities, deduced an electron/hole MPF of only ~44/17 nm [26]. But we also noted that $PMFP_{Si}$ of $0.65\mu m$ is close to the phonon MFP in silicon membrane [24], which agrees with the report that the MFP of carriers and phonons is of the same order of magnitude at high temperatures in Si [30]. Photocarriers investigated here are excited-state hot carriers occupying high positions in the energy band, typically possessing an electron temperature on the order of several thousand $K$ [31]. Our results confirm that the MFP of these high-energy photocarriers in the transient, highly non-equilibrium transport process can be significantly larger than the MFP determined electrically, which reflects only the quasi-equilibrium carriers near the Fermi level or the band edge [30, 32].

We have also characterized photocarrier transport in a GaAs quantum well, which consists of ten periods of 15$nm$ thick GaAs well and 185$nm$ thick Al$_{0.3}$Ga$_{0.7}$As barrier with a Si-$\delta$-doped layer located at the center of each barrier except the second barrier [33]. We chose a 515$nm$ pump pulse and an 850$nm$ probe pulse, the detailed experimental results are in the Supplemental Material [26]. From Fig. 4(b), we can gain $PMFP_{GaAs} \sim 1.25\ \mu m$. According to a first principal calculation [15], the MFP of electrons in GaAs in the low filed transport regime at 300 $K$, is around 130~210$nm$. The calculation is precisely applicable to describe the local-equilibrium condition in the electrical transport measurements, including the means based on the Hall effect [18], four-terminal configurations [19] and conductive atomic force microscope [20,21]. Thus, it can be inferred that the PMFP (1.25 $\mu m$) we obtained by our TFM optical method is once again much larger (10 times) than those obtained by the electrical approaches. Our findings indicate that the PMFP of Si is shorter than that of GaAs. This is primarily attributed to the fact that Si has a higher effective carrier mass [26] and an indirect bandgap structure, which enhances carrier-phonon scattering and reduces mobility. In general, materials with lower mobility tend to display weaker ballistic effects [25].

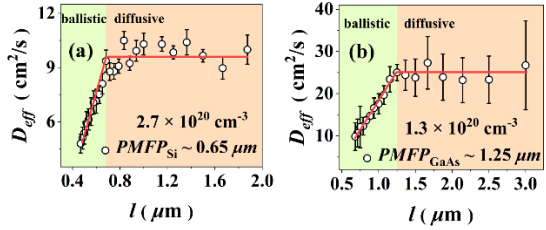

FIG 4. The effective diffusion coefficient $D_{eff}$ versus transportation characteristic length $l$ of (a) 2$\mu m$-thick Silicon membrane and (b) GaAs quantum well. The photocarriers at initial concentrations is $2.7 \times 10^{20}\ cm^{-3}$ and $1.3 \times 10^{20}\ cm^{-3}$, respectively.

Given that our experiments combine strong excitation and ultrafast detection techniques, we believe that either one of the following mechanisms or their combined action is responsible for the measured large PMFP, but conclusive evidence awaits more detailed research. i) As the relatively large single-pulse energy flux density employed, the injected carrier concentration by photoexcitation is exceedingly high (> $10^{20}$ $cm^{-3}$), which generates a Coulomb screening [34, 35] that weakens the scattering of carriers by polar phonons, resulting in a relatively long PMFP. And since the carriers are photogenerated rather than ionized, there is no carrier-charged impurity scattering in our measurements. ii) The hot electrons in a high-energy non-equilibrium state are at an elevated temperature. This thereby expands the scope and enables a greater number of large-group velocity carriers to take part in the transport process [36]. iii) The carriers generated by the 515$nm$ laser, which is much higher than the band gaps of Si and GaAs, own high kinetic energy. The phase space limitation and phonon bottleneck effect suppress the energy relaxation channel through carrier-acoustic phonons interaction and increase the scattering time [37], hence increasing the PMFP.

In summary, we proposed and demonstrated an accurate and efficient technique to measure PMFP, namely, speckle pump – TFM. The innovative design of shaping the pump light into speckles enables the measurement to cover the region from ballistic to diffusive carrier transport, and the PMFP of Silicon membrane is successfully measured as~ 0.65$\mu m$, and that of a GaAs quantum well is ~ 1.25$\mu m$. These results demonstrate that the MFP of photogenerated high-energy carriers can be an order of magnitude larger than that of local-equilibrium carriers near the Fermi surface or around the band edge, which updates the perspective on the transport behaviors of highly non-equilibrium photocarriers.

This work was supported by projects National Key Research and Development Program of China No. 2023YFB4603801; National Natural Science Foundation of China No. 52176173 and No. 21FAA02809; Guangdong Innovative and Entrepreneurial Research Team Program No. 2021ZT09L227; Guangdong Basic and Applied Basic Research Foundation No. 2020A1515110192, No. 2022A1515010710 and No. 2023B1515040023. The experiments reported were partially conducted at the Guangdong Provincial Key Laboratory of Magnetoelectric Physics and Devices. No. 2022B1212010008.

*Corresponding author: chenk35@mail.sysu.edu.cn

# Supplemental Material：
# Measurement of Photocarrier Mean Free Path via Speckled Laser Pump - Transient Fourier Microscopy Probe

Zuni Luo,[1,2] Yifan Zhang[1,2], Jie Wei[2], Zhikun Xie[2], Tianshu Lai[2], and Ke Chen[1,2, *]

[1]Center for Neutron Science and Technology, Guangdong Provincial Key Laboratory of Magnetoelectric Physics and Devices, School of Physics, Sun Yat-sen University, 510275 Guangzhou, China

[2] State Key Laboratory of Optoelectronic Materials and Technologies, School of Physics, Sun Yat-sen University, 510275 Guangzhou, China

[*]Corresponding author: chenk35@mail.sysu.edu.cn

**Experimental Setup：**

The experimental setup is adding a reference light and a shaping module to the ultrafast pump-probe reflectivity microscopy, as shown in Fig. 1(a). A seed laser (1030*nm*, < 190 *fs*, 200 *kHz*, 100 *µJ* per pulse, PHAROS, Light Conversion) was used to pump an optical parametric amplifier (OPA, ORPHEUSF, Light Conversion) to generate excitation or detection pulses. The 515*nm* pump beam generated by the frequency-doubling crystal directly output from the OPA, while the 650*nm* (for Silicon membrane) and 850*nm* (for GaAs quantum well) acting as the probe beam export after a simple prism-based compressor (< 55 *fs*). The pump beam is collimated into the imaging objective (MPAL, 50×, Cossim), while the probe beam passing through the delay line (ILS200, Newport) is pre-focused by the lens L1 ($f$ = 1000 *mm*) before entering the objective to achieve widefield detection. The reflected probe beam carrying the carrier relaxation dynamics information from the sample is collected by the objective and focused by the tube lens L2 ($f$ = 200 *mm*) onto a CCD (ME2P - 2622 - 15U3M NIR, Daheng Imaging) for imaging. A fraction of the modulated pump beam is led out by the beam splitter BS1, then reflecting to the edge of view to act as the reference light, as shown in Fig. 1(b). The rectangular imaging area needs to be fully illuminated by the probe beam and the reference beam should be located in the corner where is far away from the reflection image signal. The pump beam is located in the center of the square image, which is obtained by cropping the reference beam late and used as the raw data for TFM analysis. A diffuser is placed at the focal plane of two objectives (HK, 60×, Hengyang) to shape the pump beam into speckles with an average size of 423*nm*, as shown in Fig. S1(a) and Fig. S1(b). In our experiment, the image size is set to 600×600 pixels.

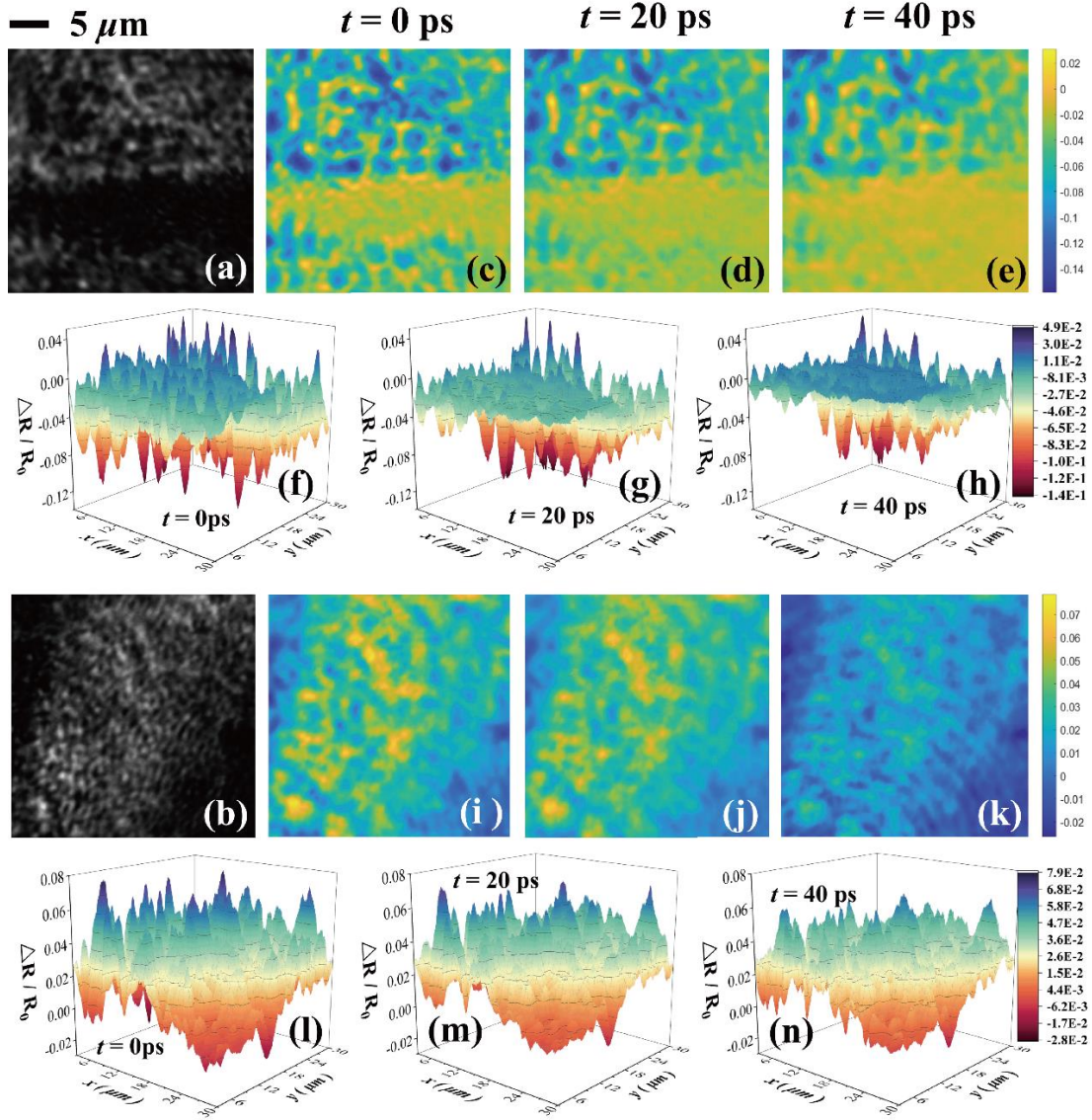

FIG S1. The distribution of photogenerated carriers at different times under speckle pump. (a) and (b) are the morphologies of the speckle pump on the 2$\mu m$-thick Silicon membrane and GaAs quantum well, respectively. (c)-(e) display the Si reflection image signals at $t = 0\ ps$, $t = 20\ ps$ and $t = 40\ ps$. Since the electron signal of Si is negative, which means the photogenerated carriers cause a decrease in reflectivity, the negative values shown in blue represent the carrier distribution. (f)-(g) are the corresponding 3D views, where the red representing negative values indicates the distribution of photogenerated carriers at the different moment. (i)-(k) represent the diffusion of carriers in the GaAs quantum well at three different delay times. The electron signal of the GaAs quantum well is positive, so the carrier distribution corresponds to the yellow positive regions. The three-dimensional views (l)-(n) show the distribution of photocarriers at the corresponding time, represented by positive values in blue. Obviously, the morphology of the speckle pump corresponds one-to-one with the distribution of the photogenerated carriers. As time passes, the carriers in the two materials gradually diffuse outward and also disappear due to recombination. All the above pictures share the same length scale.

**Virtual lock-in amplification technique:**

The signal input is obtained by the area-array camera in the microscopic imaging system. After taking a series of pictures under the fixed delay line, then extracting the intensity change of each pixel over time, multiple pump-probe signals with spatial distribution characteristics can be collected at once. Considering that the reflection image signal excited by pump beam will change with the signal variation, the phase interference of noise cannot be ignored. As for reference acquisition, we adopt the scheme of extracting the pump beam to ensure that the reference light is strong enough for recording the modulated of the pump beam in real-time. It maintains the same frequency and phase as the input signal, which can effectively avoid the influence of system errors such as unstable frequency of the optical modulator and frame dropping. Notably, the imaging position of the reference light should be far away from the area acted by the pump beam to avoid its interference with the signal. Here, we set the camera frame rate to the maximum value to ensure the modulation effect of chopper can be distinguished, and the exposure time is 6 *ms* which makes the pump laser induced reflection averaged for 1200 times in single frame for effectively enhancing the reflection signal and suppressing the background noise. At the same time, the shooting duration is controlled at ten thousand frames, which can be equivalently approved that the input is an approximately infinite-time periodic signal required in the relevant calculation integral formula. Phase-sensitive detection is to perform a correlation operation on these two signals, so that the carrier distribution at a certain moment can be visualized through the pump-probe signals with spatial distribution characteristics,

$$n(x,y,t) = \frac{\Delta R(x,y,t)}{R_0(x,y,t)} = \frac{\int_{t=0}^{t=T} I_{pump}(x,y,t) \cdot I_1(x_1,y_1,t)dt}{\int_{t=0}^{t=T} I_{probe}(x,y,t) \cdot I_2(x_2,y_2,t)dt} \tag{S1}$$

where $n(x,y,t)$ refers to the carrier distribution at time *t* in the coordinates $(x,y)$. After modulating the pump beam by chopper, $I_{pump}(x,y,t)$ records the grayscale of the image taken at time *t* in the position coordinates $(x,y)$, $I_1(x_1,y_1,t)$ represents the grayscale of a certain position within the reference light area of the same picture at time *t* in the coordinates $(x_1,y_1)$, which is the reference light reflected to the edge of view to reflect the process of the modulated pumping beam in time. The coordinates $(x_1,y_1)$ can be randomly selected within the space where the reference light is recorded. Signal $\Delta R(x,y,t)$ is the result of the relevant operation by these two items within the time period from $t=0$ to $t=T$. $I_{probe}(x,y,t)$ represents the grayscale value in the position coordinates (x,y) of the picture taken at time *t* after blocking the pump beam and modulating the probe beam, $I_2(x_2,y_2,t)$ is the grayscale in a fixed coordinate $(x_2,y_2)$ of the image taken at time *t*. This coordinate can be arbitrarily selected in the imaging area which is used to record the change of the probe beam with time. Similarly, the relevant operation of these two terms within the time period from $t=0$ to $t=T$ can obtain the background $R_0(x,y,t)$. To prevent the introduction of false dynamic signals resulting from the non-strict collimation of the delay line, it is advisable to acquire a series of video sequences respectively for both signal and background at each time interval. As evident from Equ. (S1), the probe beam must be fully illuminated to ensure complete coverage of the imaging area, thereby avoiding a denominator value of zero. Specifically, as the reference signal $I_1(x_1,y_1,t)$ and $I_2(x_2,y_2,t)$ are normalized which the DC component of the amplitude is removed, while the gray values $\geq 0$ are set to 1 and those below zero are assigned a value of -1.

Fig. S2(a) and S2(b) show the period of the signal $\Delta R(x,y,t)$ and the background $R_0(x,y,t)$, respectively. In Fig. S2(a), the red dashed line denotes the reference light $I_1$ falling on the edge of view, while the black solid line is the probe beam signal $I_{pump}$ after modulating the pump beam with a chopper of $f = 10\ Hz$. Clearly, the square wave displayed by $I_1$ owns different widths with slight deviations, implying the frequency instability of the chopper is records continuously. The signal $I_{pump}$ always remains synchronized with $I_1$, although appearing rough with lots of system noise. Fig. S2(b) demonstrates the background $R_0(x,y,t)$ of the center $I_2$ (red dashed line) and the corner $I_{probe}$ (black solid line) of the captured image after modulating the probe beam when the pump beam stops, which also reflects that the modulation frequency of the chopper is unstable and the signals are highly isochronous. In our experiment, a video includes 715 such cycles. For the convenience of comparative observation, the amplitudes of $I_1$ and $I_2$ have been amplified to be similar to those of $I_{pump}$ and $I_{probe}$ respectively, while in the actual algorithm, $I_1$ and $I_2$ are normalized.

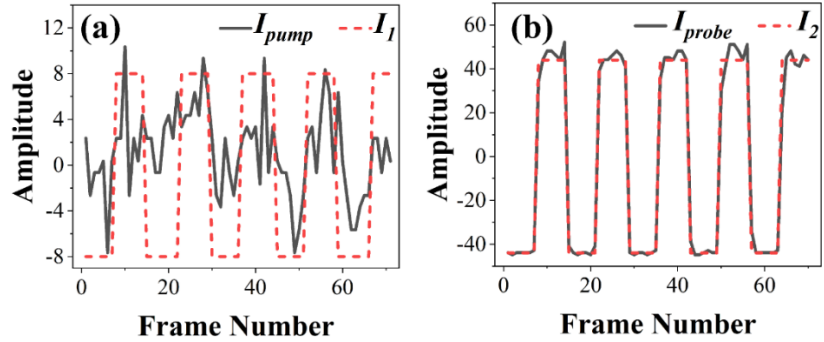

FIG S2. The principle of the virtual lock-in amplification technology. (a) and (b) are the periodogram of the signal $\Delta R(x,y,t)$ and the background $R_0(x,y,t)$ respectively, show the results of local magnification, displaying only 5 cycles.

Fig. 1(c) and 1(d) exhibit the reflection image signal $\Delta R/R_0$ at $t = 18$ ps when pre-use and post-usage the virtual lock-in amplification technology. The pump light is Gaussian-shaped and the electronic signal of GaAs is positive, which means the photogenerated carriers cause an increase in reflectivity, hence the distribution of carriers at different times is manifested as a positive-peak Gaussian shape, corresponding to the middle region of the three-dimensional image. In Fig. 1(c), since the carriers are accompanied by recombination while diffusing, the signal amplitude decrease and even drop below the background noise, which generates the SNR is too low to observe the diffusion trend. After employing the virtual lock-in amplification technology, this problem can be effectively improved, as shown in Fig. 1(d), where the SNR has enhanced from 2.2 to 16.5. This result resoundingly confirms that the proposed virtual lock-in amplification technology can be used to detect weak signals submerged in noise by increasing the SNR more than seven times, effectively realizing weak energy density pump measurement and more accurately visualizing the carrier transportation on an ultrafast time scale. The SNR in Fig. 1 (c) and Fig. 1(d) are calculated by the noise standard deviation method, as shown in Fig. S3.

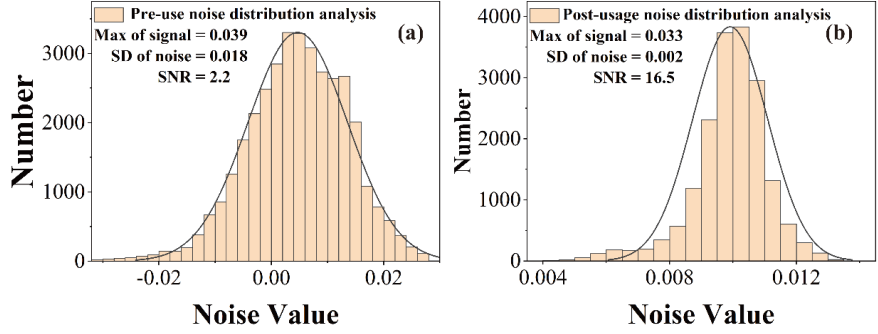

FIG S3. The SNR of $\Delta R/R_0$ before and after using the virtual lock-in amplification technique at $t = 18$ ps. The SNR is obtained by the noise standard deviation method, which is sampling each noise signal and statistically presenting as a histogram, then fitting the distribution by Gaussian, and the full width half maximum obtained is the noise standard deviation. Specifically, (a) explains the noise distribution of Fig. 1(c), while (b) corresponds to Fig. 1(d).

**Theory for the size effect between ballistic and diffusion transport:**

In nano-devices, when the electron transport length $l$ is shorter than or nearly equal to the mean free path $MFP$, the conductance is measured as a constant, which is designated as ballistic conductance $G_B$. It can be related to the conductance $G$ by

$$G = \frac{G_B \cdot MFP}{l + MFP} \tag{S2}$$

The relationship between conductance $G$ and effective conductivity $\sigma_{eff}$ is expressed by

$$G = \frac{\sigma_{eff} \cdot A}{l} \tag{S3}$$

where $A$ denotes the cross-sectional area of the conductor. In the realm of non-degenerate semiconductors, the carriers adhere to the Boltzmann statistics, whose effective conductivity $\sigma_{eff}$ satisfies:

$$\sigma_{eff} = \frac{e^2 n_c}{k_B T} \cdot D_{eff} \tag{S4}$$

where the Boltzmann constant $k_B$, temperature $T$, electron charge $e$ and carrier density $n_c$ are fixed in the same system. By integrating equations (S2) - (S4), the effective diffusion coefficient $D_{eff}$ is given by

$$D_{eff} = \frac{k_B T G_B \cdot MFP}{e^2 n_c A} \cdot \frac{l}{l + MFP} \tag{S5}$$

Since these parameters remain are fixed in the same system, their product can be substituted with a constant. As for the degenerate semiconductor, given that the Fermi level lies deep within the energy band, the carriers follow Fermi-Dirac statistics. Under this circumstance, equation (S4) is adjusted as:

$$\sigma_{eff} = e^2 \cdot \frac{\partial n_c}{\partial E_F} \cdot D_{eff} \tag{S6}$$

where $\frac{\partial n_c}{\partial E_F}$ represents the density of states filling factor, which is also a constant term. Through theoretical derivation, the effective carrier diffusion coefficient in any semiconductor can be deduced expressed as:

$$D_{eff} = C \cdot \frac{l}{l+MFP} = \begin{cases} C \cdot l/MFP & (l \ll MFP) \\ C & (l \gg MFP) \end{cases} \quad (S7)$$

where $C$ is the diffusion coefficient of diffusion zone. Equation (S7) effectively depicts the transition of the carrier transport state from ballistic to diffusion.

Based on the size effect, when the transport length $l$ is relatively large, the contribution of the mean free path $MFP$ in the denominator becomes negligible. In this case, $D_{eff} \approx C$ remains constant, signifying that the carriers are in the diffusion regime. Conversely, when $l \ll$ MFP, the term $l$ in the denominator can be disregarded, resulting in $D_{eff} \approx C \cdot l/MFP$, which is proportional to $l$, indicating that the carriers are in the ballistic regime. It should be emphasized that the term $l$ in the formula represents the transportation characteristic length of the system. For a device with voltage applied, $l$ corresponds to the distance from the source to the drain. In the case of our system, the $l$ corresponds to half a grating period because the electron source can be regarded as the peak, and the electron drain as the valley of the grating structure.

## The electron/hole MPF of 2*μm*-thick Silicon membrane gained by four-probe method:

The four-probe method involves simultaneously pressing four metal probes with a spacing of 1*mm* onto the flat surface of the measured sample. A constant current source is used to pass a small current $I$ through probes 1 and 4, and then the voltage $V$ is measured across probes 2 and 3. The resistivity $\rho$ of the sample is given by

$$\rho = C_1 \frac{V}{I} \quad (S8)$$

where $C_1$ is the correction coefficient of the four-probe setup, which is a constant when the probe positions and spacing are determined. Using this method, the resistivity of 2*μm*-thick Silicon membrane was measured to be 21.9 Ω·cm. The conductivity $\sigma$, which is the reciprocal of the resistivity $\rho$, satisfies the relationship:

$$\sigma = \frac{1}{\rho} = n_c e \mu \quad (S9)$$

where $\mu$ is the mobility. By substituting the reference value 1400 $cm^2 \cdot V^{-1} \cdot s^{-1}$ for intrinsic Si, the corresponding carrier concentration $n_c = 2.04 \times 10^{14}\ cm^{-3}$. For this non-degenerate semiconductor, the electron MFP is expressed by

$$MFP = \mu \frac{m^*}{e} \cdot \sqrt{\frac{8k_B T}{\pi m^*}} \quad (S10)$$

Substituting the effective mass of electron/hole $m^* = 0.26/0.36\ m_0$ (where $m_0$ is the electron rest mass), the Boltzmann constant $k_B$ and the measurement temperature $T = 300\ K$, the electron/hole MFP of 2*μm*-thick Silicon membrane is calculated to be ~ 44/17 *nm*.

TABLE S1. The effective mass of electron/hole in Si and GaAs

| Sample | Electron | Light Hole | Heavy Hole | Density of States for Holes |
|---|---|---|---|---|
| Si | $0.26\ m_0$ | $0.16\ m_0$ | $0.49\ m_0$ | $0.36\ m_0$ |
| GaAs | $0.063\ m_0$ | $0.082\ m_0$ | $0.48\ m_0$ | $0.33\ m_0$ |

## Experimental results of GaAs quantum well:

In Fig. S4(a) and S4(b), the normalized amplitude spectra at different wave vectors decay faster than those of Silicon membrane, indicating more rapid carrier diffusion. In Fig. S4(c), at the diffusion region $q = 0.8 \sim 2.5\ \mu m^{-1}$, the bipolar diffusion coefficients gained under the Gaussian pump mode $D_{G-GaAs}$ ($24.9 \pm 3.3\ cm^2/s$) and the speckled pump mode $D_{S-GaAs}$ ($25.7 \pm 1.0\ cm^2/s$), are close to the $D_{TRM-GaAs}$ ($25.1 \pm 0.7\ cm^2/s$) obtained by the transient reflectivity microscopy (see Fig. S4(d)). The measured values are consistent with the literature one [1]. Meanwhile, the carrier lifetimes $\tau_{G-GaAs}$ ($2.3 \pm 1.6\ ns$) and $\tau_{S-GaAs}$ ($2.4 \pm 1.1\ ns$) are also almost consistent with $\tau_{PP-GaAs}$ ($2.3 \pm 0.3\ ns$) measured by the pump-probe technique (see Fig. S6(b)). The speckle pump can generate a large range of effective wave vectors with $q_{max} = 4.6\ \mu m^{-1}$. And according to the observation that the experimental decay rate in Fig. S4(c) shows obvious deviation from the diffusive slope again, we can infer that the system has successfully arrived the ballistic transport region for large $q$. In Fig. S4(d), due to the use of the virtual lock-in amplification technology, the improvement of SNR enables successful fitting of the carrier wave packet even at a late delay time $t = 2300\ ps$ when there are very few surviving photocarriers remaining, which greatly highlights the superiority and necessity of this technology. The excited carrier density range ($1.0 \sim 4.9 \times 10^{20}\ cm^{-3}$) of the above experiments are all within the linear regions characterized by Fig. S7(c) and Fig. S7(d).

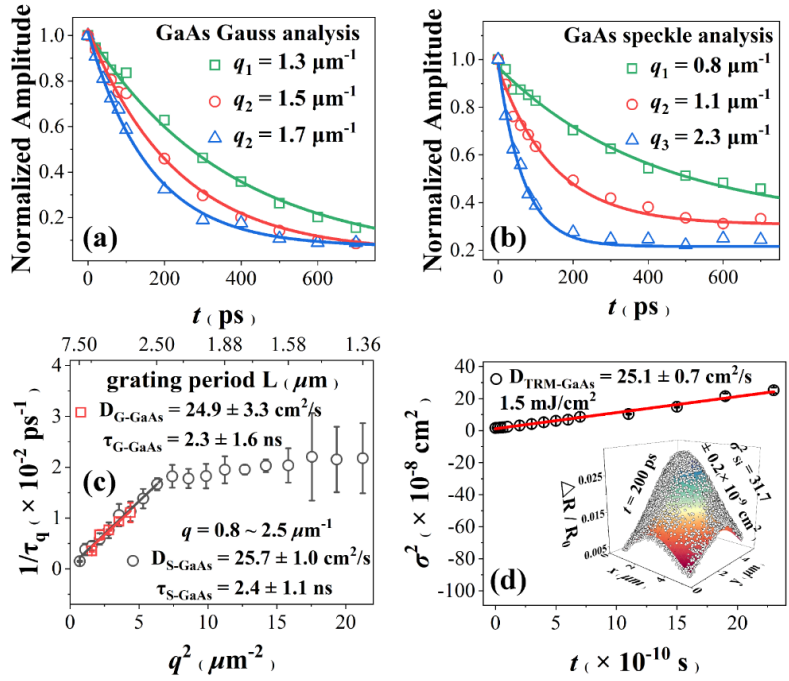

FIG S4. The experimental results of GaAs quantum well. (a)-(c) are the results of TFM. (d) The result of transient reflectivity microscopy obtained from the same set of original data. The inlaid Figure is the fitting result of the 2D Gaussian profile at $t = 200\ ps$.

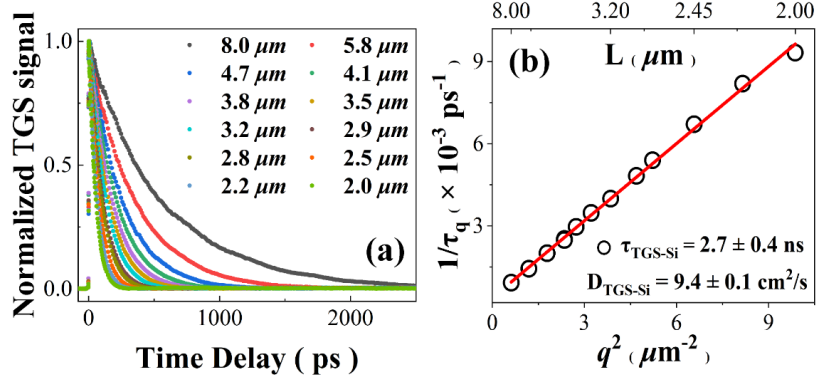

FIG S5. Transient grating experiments for 2μm-thick Silicon membrane. (a) The evolution of the normalized TGS signal under different grating periods. (b) The grating decay rate versus wave vector squared showing the diffusive behavior of carriers.

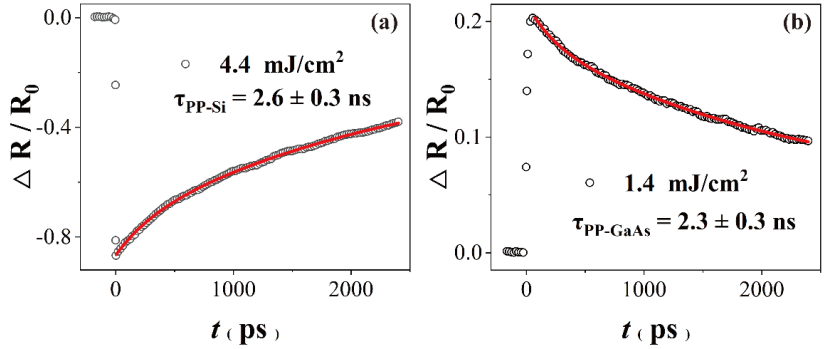

FIG S6. The reflection signal measured by pump-probe technique. Here, we place the long-focus lens L1 into the pump beam path, making the pump spot larger than the probe in the imaging area to ensure that the probe light only carries the recombination information of carriers. By performing exponential fit on the signal, we obtain: (a) the carrier lifetime $\tau_{PP-Si} = 2.6 \pm 0.3\ ns$ of the 2μm-thick Silicon membrane at an energy flux density 4.4 $mJ/cm^2$, (b) $\tau_{PP-GaAs} = 2.32 \pm 0.26\ ns$ of the GaAs quantum well under an energy flux density 1.4 $mJ/cm^2$.

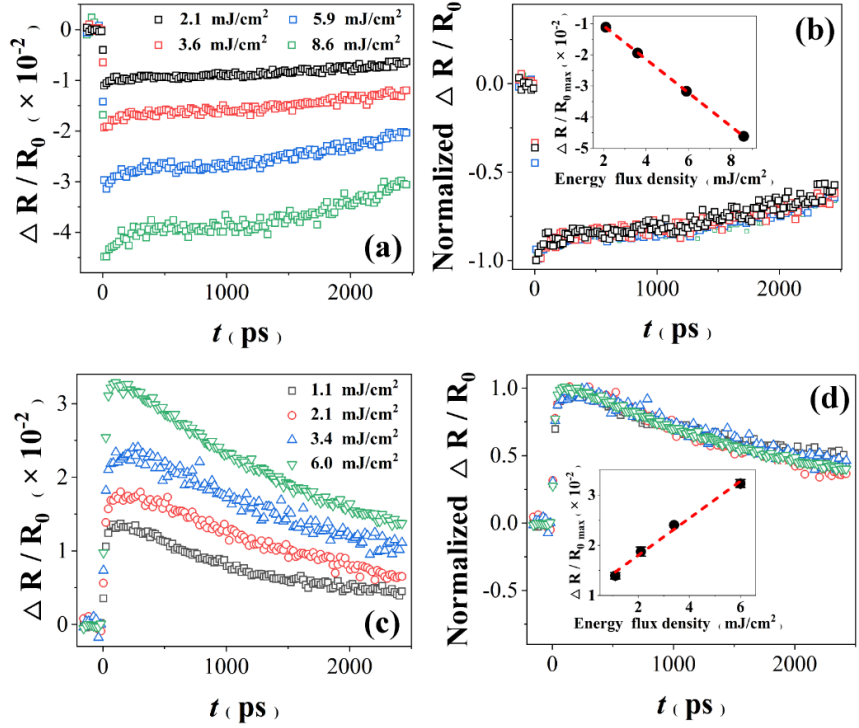

FIG S7. The pump-probe reflection signals of Gaussian pump under four energy flux densities. The reflection signals in (a) for 2$\mu m$-thick Silicon membrane and (c) for the GaAs quantum well are collected when the pump spot is smaller than the probe under Gaussian pump. After normalization, the curves in (b) and (d) overlap, and the extreme values satisfy linearity, indicating that the experimental systems are in the linear range when the energy flux density is 2.1 ~ 8.6 $mJ/cm^2$ for Si and that is 1.1 ~ 6.0 $mJ/cm^2$ for GaAs.

---

*Corresponding author: chenk35@mail.sysu.edu.cn*Corresponding author: chenk35@mail.sysu.edu.cn